\documentclass[]{spie}  

\usepackage{amsmath,amsfonts,amssymb}
\usepackage{graphicx}
\usepackage[colorlinks=true, allcolors=blue]{hyperref}

\title{The Infra-Red Telescope (IRT) on board the THESEUS mission}

\author[a]{Diego G\"otz}
\author[b]{Stéphane Basa}
\author[a]{Frédéric Pinsard}
\author[b]{Laurent Martin}
\author[a]{Axel Arhancet}
\author[c]{Enrico Bozzo}
\author[a]{Christophe Cara}
\author[e]{Isabel Escudero Sanz}
\author[a]{Pierre-Antoine Frugier}
\author[b]{Johan Floriot}
\author[c]{Ludovic Genolet}
\author[d]{Paul Heddermann}
\author[a]{Emeric Le Floc'h}
\author[a]{Isabelle Le Mer}
\author[c]{Stéphane Paltani}
\author[b]{Tony Pamplona}
\author[b]{Céline Paries}
\author[e]{Thibaut Prod'homme}
\author[a]{Benjamin Schneider}
\author[d]{Christoph Tenzer}
\author[a]{Thierry Tourrette}
\author[a]{Henri Triou}

\affil[a]{AIM, CEA-Irfu/DAp, CNRS, Université Paris-Saclay, F-91191 Gif-sur-Yvette, France }
\affil[b]{Laboratoire  d'Astrophysique  de  Marseille,  UMR  7326,CNRS,  Universit\'e  d'Aix  Marseille,  38,  rue  Fr\'ed\'eric  Joliot-Curie,Marseille, France}
\affil[c]{Department of Astronomy, University of Geneva, Chemin d’Ecogia 16, 1290 Versoix, Switzerland}
\affil[d]{Institut für Astronomie und Astrophysik, Abteilung Astronomie, Universität Tübingen, Sand 1, D-72076 Tübingen, Germany}
\affil[e]{European Space Agency, ESTEC, Keplerlaan 1, 2201 AZ, Noordwijk, The Netherlands}	

\authorinfo{ E-mail: diego.gotz@cea.fr}

\begin{document} 
\maketitle

\begin{abstract}
The Infra-Red Telescope (IRT) is part of the payload of the THESEUS mission, which is one of the two ESA M5 candidates within the Cosmic Vision program, planned for launch in 2032. The THESEUS payload, composed by two high energy wide field monitors (SXI and XGIS) and a near infra-red telescope (IRT), is optimized to detect, localize and characterize Gamma-Ray Bursts and other high-energy transients. The main goal of the IRT is to identify and precisely localize the NIR counterparts of the high-energy sources and to  measure their distance. Here we present the design of the IRT and its expected performance.

\end{abstract}

\keywords{Gamma-Ray Bursts, infra-red telescopes and instrumentation}

\section{INTRODUCTION}
\label{sec:intro}  

THESEUS (Transient High Energy Sky and Early Universe Surveyor) is one of the candidate M5 missions within the ESA Cosmic Vision program (see Amati et al. this volume). It is currently in phase A (until mid 2021), and if THESEUS is selected, it will be launched 2032. 
The THESEUS mission is being designed as a multi-wavelength observatory, whose scientific goal is to increase the discovery space of the high-energy transient phenomena and especially Gamma-Ray Bursts over the entirety of cosmic history, and especially at redshifts larger than 5.5. The payload to be carried by THESEUS is composed by the XGIS (X and Gamma-ray Imaging Spectrometer) coded mask telescopes, operating in the 2 keV--10 MeV energy range, the SXI (Soft X-ray Imager) focusing telescopes, operating in the 0.3--5 keV energy range and the near-Infra-Red Telescope (IRT) sensitive in the 0.7--1.8 micron band. 
The main goal of the IRT is to identify, accurately localize ($\leq$1$''$) and measure the distance of  the NIR counterparts (redshift determination to an accuracy of 10\% or less) of the high-energy sources discovered by the XGIS and SXI. 
In addition, the IRT will be used to characterize the afterglows, through spectroscopy for a part of them (spectroscopic redshift, neutral hydrogen absorption, presence of metals), and as a multi-purpose agile NIR observatory in space through the implementation of a dedicated Guest Observer (GO) programme, and a Target of Opportunity programme, with special emphasis on multi-messenger and time-domain astrophysics. 

The IRT has thus been designed in order to implement imaging capabilities over a wide field of view (15$^\prime$ $\times$ 15$^\prime$), and moderate resolution spectroscopy (R$\sim$400) over a smaller portion of the field of view (2$^\prime$ $\times$ 2$^\prime$).  The IRT responsibility is shared among ESA (telescope, thermal control, and detector procurement) and a consortium lead by France, in collaboration with Switzerland and Germany, that will deliver the IRT instrument.

\section{The IRT Telescope}

The telescope optical concept is under responsibility of ESA to guarantee any opto-thermo-mechanical concepts developed by the different prime contractors can interface with the single design of the IRT instrument. 

\begin{figure} [ht]
   \begin{center}
   \begin{tabular}{c} 
   \includegraphics[height=7cm]{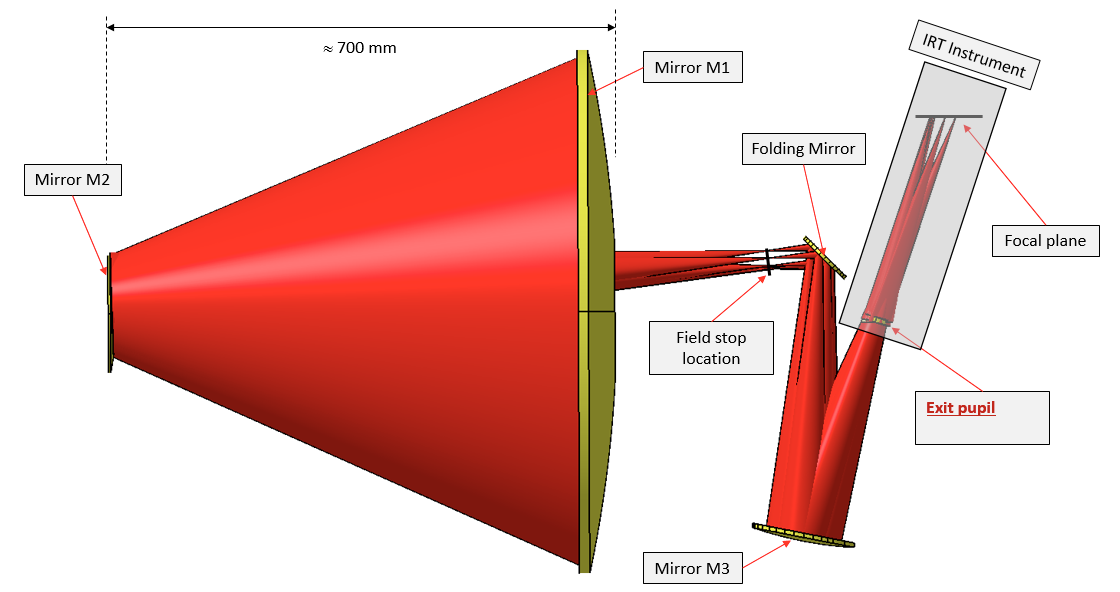}
   \end{tabular}
   \end{center}
   \caption[example] 
   { \label{fig:fov} 
IRT Optical scheme. M1, M2, M3 are under ESA/Prime responsibility. The exit pupil represents the optical interface with the IRT Instrument provided by the consortium.}
   \end{figure}

The IRT telescope is a focusing three-mirror Korsch configuration with an off-axis (0.884$^\circ$) field. Between the secondary and the tertiary mirrors of the Korsch, there are two additional optical components: a field stop at the intermediate focus and a flat folding mirror, see Fig. \ref{fig:fov}, to facilitate accommodation in the spacecraft. Note that in the absence of the at folding mirror, the three powered mirrors of the Korsch (M1, M2
and M3) are coaxial, the common axis being coincident with the axis of radial symmetry of each mirror. The mirror shapes are conics of revolution. The aperture stop is at the primary mirror. Its image, the exit pupil, in the converging output beam is the interface with the IRT instrument. Table \ref{tab:telescope} gives a summary of the telescope key figures.

\begin{table}[ht]
\caption{Main characteristics of the IRT telescope.} 
\label{tab:telescope}
\begin{center}       
\begin{tabular}{|c|c|} 
\hline
Telescope type & Off-axis Korsch  \\
\hline
Entrance pupil diameter & 700 mm\\
\hline
M1-M2 distance & 675 mm\\
\hline
Exit pupil diameter & 36 mm\\
\hline
Collecting area & $>$ 0.34 m$^2$\\
\hline
Wavelength range & 700--1800 nm \\
\hline
 Throughput & $>$ 80 \%\\
 \hline
 Pixel scale & 0.6 arc sec \\
 \hline
 Focal length & 6188 mm\\
 \hline
 
\end{tabular}
\end{center}
\end{table}

The optical scheme of the telescope is sketched in Figure \ref{fig:fov}. The optical design will implement two separated field of views, one for photometry with a minimal size of 15 $\times$ 15 arc min (potentially extended to 17 $\times$ 20 arc min), and one for spectroscopy of 2 $\times$ 2 arc min, see Figure \ref{fig:fov}. On the photometric field of view the IRT will be able to acquire images using five different filters (I, Z, Y, J and H) and on the spectroscopic field of view the IRT will provide moderate resolution (R$\sim$400 at 1.1 $\mu$m) slit-less spectroscopy in the 0.8-1.6 microns range.

\begin{figure} [ht]
   \begin{center}
   \begin{tabular}{c} 
   \includegraphics[height=6.5cm]{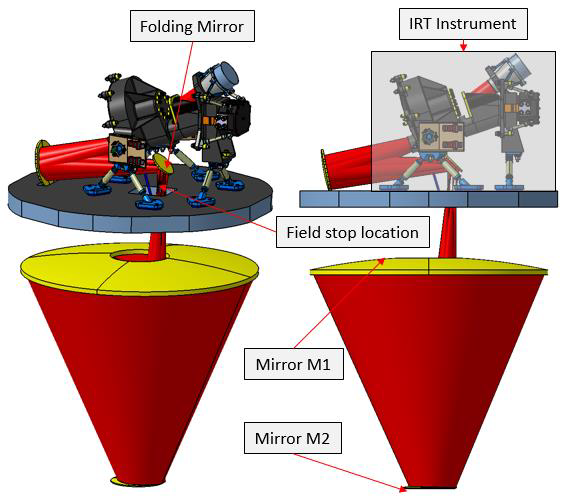}
      \includegraphics[height=6.5cm]{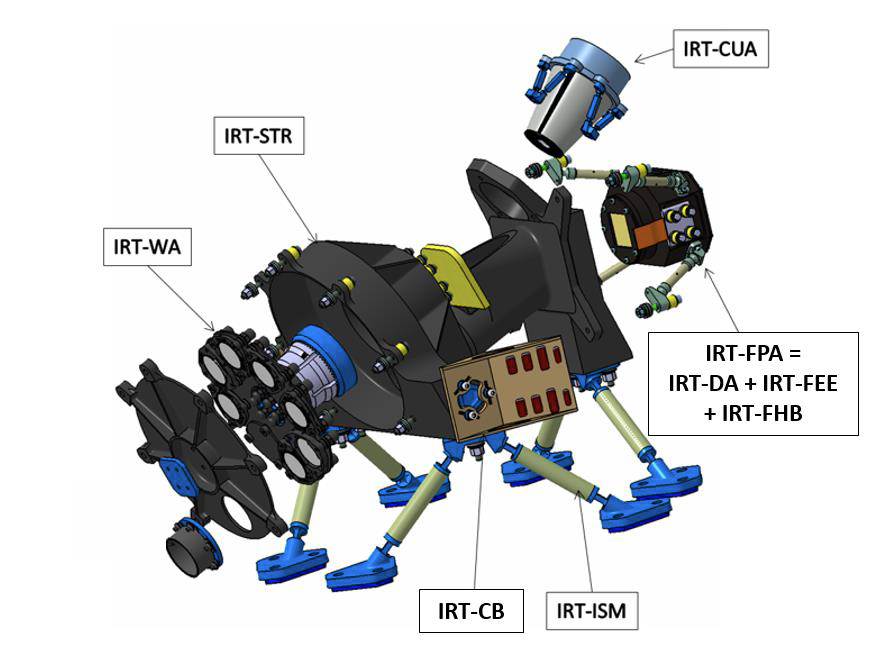}
   \end{tabular}
   \end{center}
   \caption[example] 
   { \label{fig:view} 
Left: Full optical scheme of the IRT, including IRT-CAM and optical bench. Right: exploded view of the IRT-CAM and its subsystems (without insulating blankets).}
   \end{figure}

The different observation modes will be implemented through the design of the IRT Camera (IRT--CAM) that includes a filter wheel, carrying the different optical filter, as well as a grism, which will allow for spectroscopy, see Fig. \ref{fig:view}. The IRT-CAM will consist mainly in a structure (IRT--STR), an isostatic mount (IRT-ISM), a filter wheel assembly (IRT--WA, see section \ref{sec:wa}), a calibration unit assembly (IRT--CUA), and a focal plane assembly (IRT--FPA), hosting the detector and its associated cold electronics. The whole structure will be covered with thermal isolating blankets.
The detector currently envisaged for the IRT FPA is a Teledyne H2RG, sensitive in the 0.7--2.5 microns wavelength range.

\subsection{The Wheel Assembly}
\label{sec:wa}

\begin{figure} [ht]
   \begin{center}
   \begin{tabular}{c} 
   \hspace{-1cm}
   \includegraphics[height=6.5cm]{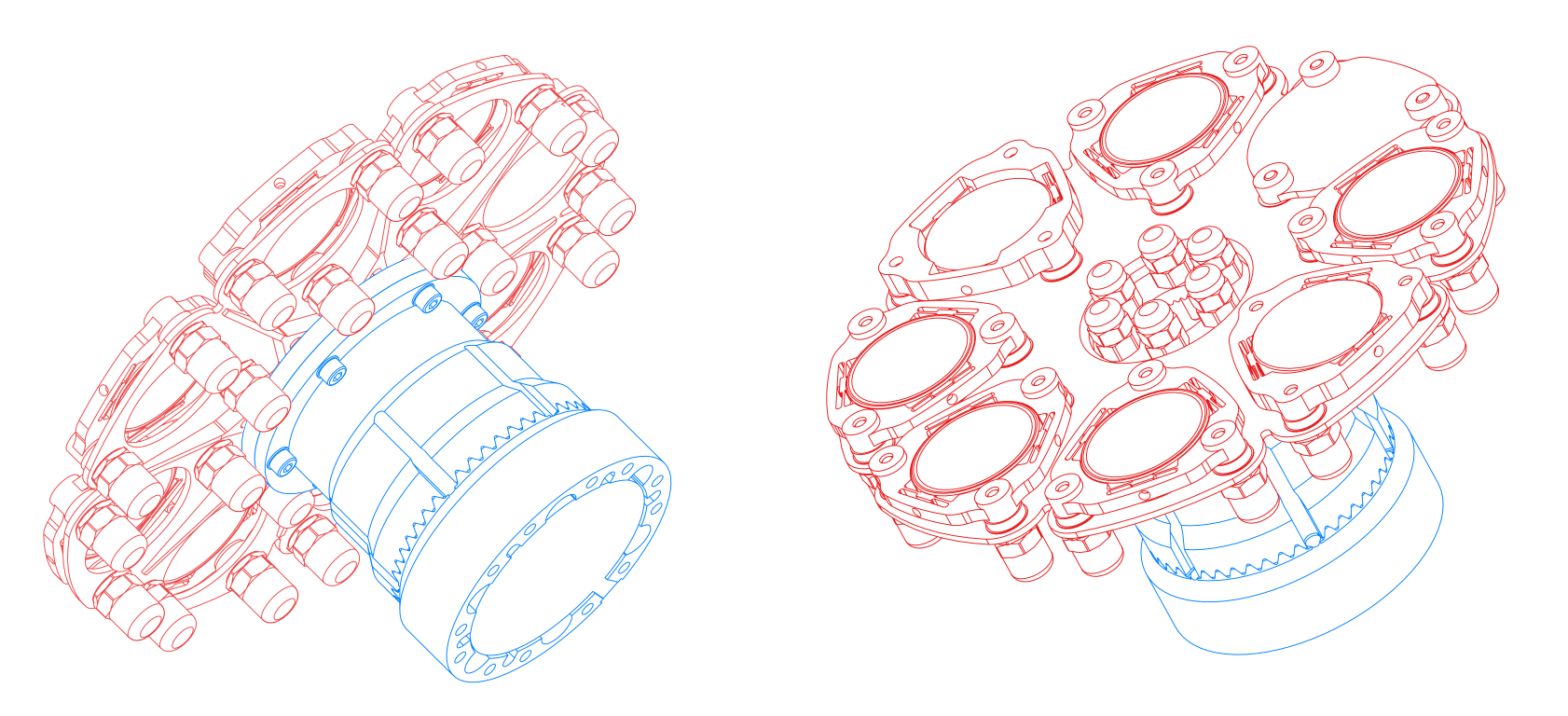}
   \end{tabular}
   \end{center}
   \caption[example] 
   { \label{fig:wheel} 
Filter wheel assembly. Actuator assembly in blue and wheel structure in red.}
   \end{figure}

The IRT wheel assembly is a rotating mechanism that ensures the positioning in the beam of the various optical components. The mechanism has eight positions (one grism, five filters, one open position (window) and one closed position). It consists of two sub-assemblies: the wheel structure and the actuator assembly. These are sketched in Figure \ref{fig:wheel}. The wheel is made of invar to ensure an excellent mechanical and thermal stability of the optical components in the whole IRT temperature range. Invar is also used for the bottom part of the actuator assembly due to constraints at the mechanism mounting interface on the IRT--STR. The actuator assembly comprises a SAGEM stepper motor, the shaft, and the angular ball bearings. The shaft is made of stainless steel 15-5PH for a good thermo-elastic compatibility with the ball bearings and the motor rotor. A hirth coupling, using tapered teeth oriented towards a central point, is implemented between parts made from different materials (invar and stainless steel). The hirth coupling teeth mesh together, stiffening the mechanism and allowing relative radial motions between the invar and stainless steel parts due to the CTE mismatch. This ensures that the angular position between these parts is kept within the required tolerances in the range of temperatures of interest for the IRT operations. An Eddy current contact-less positional sensor is also included in the actuator assembly in order to provide an angular reference position. 

\section{Interfaces and resources}

The optical interface of the IRT--CAM and the telescope is at the telescope exit pupil corresponding to the filters positions in Fig. \ref{fig:view}. The IRT instrument consists of the IRT--CAM and of the IRT Data Handling Unit (IRT--DHU). Both the IRT--CAM and the IRT--DHU will be provided by the Instrument Consortium to the Prime Contractor through ESA as Customer Furnished Items (CFIs). The integration of the IRT--CAM and IRT--DHU with the IRT telescope onto the spacecraft will be the responsibility of the Prime contractor. The IRT--CAM is accommodated onto the IRT main bench that also supports the telescope.

\begin{figure} [ht]
   \begin{center}
   \begin{tabular}{c} 
   \includegraphics[height=6.5cm]{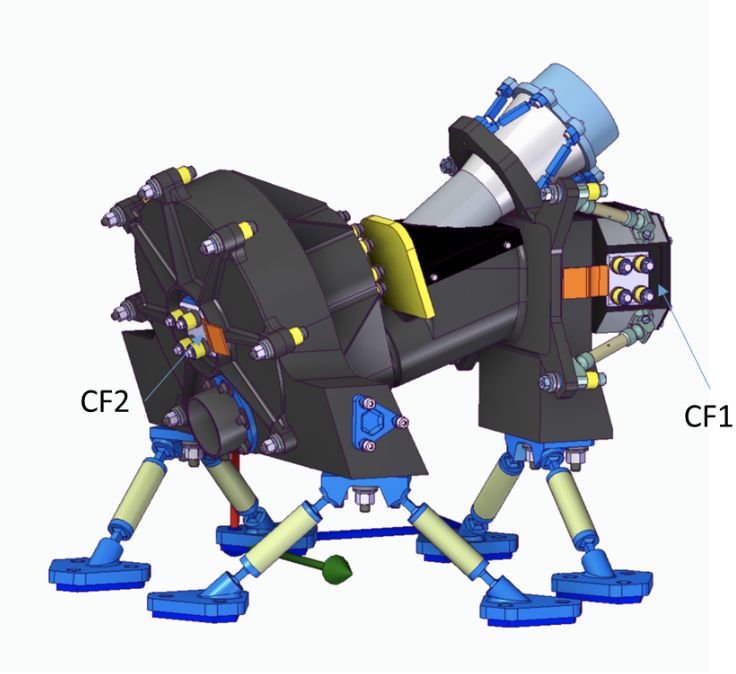}
      \includegraphics[height=6.5cm]{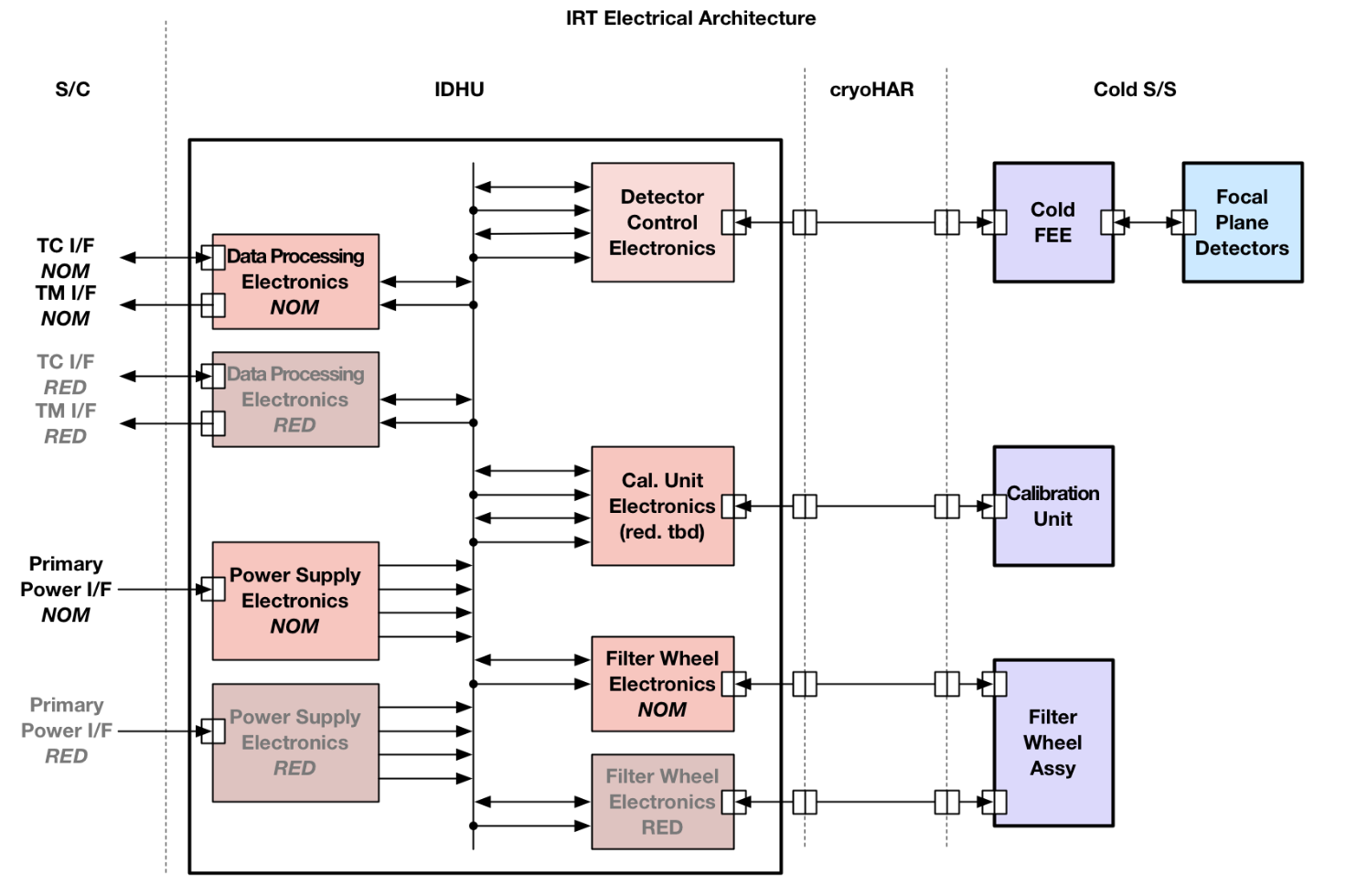}
   \end{tabular}
   \end{center}
   \caption[example] 
   { \label{fig:cf} 
Left: Representation of the IRT instrument cold fingers (orange colour). Right: IRT Instrument electrical architecture.}
   \end{figure}

The cooling of the IRT Camera is ensured thanks to two Cold Fingers (CF1 and CF2) connected to a cryo-cooling system (the cryo-cooler(s) will be provided by the Prime Contractor and the thermal interface to the instrument consists in the cold fingers). Figure \ref{fig:cf} shows the intended position for CF1 and CF2: CF2 will be on the IRT--WA and CF1 on the IRT--FPA.

Taking into account the scientific requirements in terms of thermal background, we defined the following temperatures for the IRT-CAM: the ISA and the WA shall be kept at 160K (CF2), while the detector assembly (DA) and the FPA shall be kept at 120K (CF1), thanks to the cryo-cooler(s).

To maintain the temperature stability of the camera (at IRT--FEE (Front-End-Electronics), IRT--FPA, and IRT--WA levels), we will consider, at instrument level, to use temperature probes as well as heaters that shall be controlled by a dedicated electronic (function within the Detector Control Electronics IRT--DCU and Filter Wheel control electronics IRT-WCU (TBC)). As a baseline, we consider that the cryocooler will provide the specified temperature and thermal power at the cold finger (CF1 and CF2) and that the fine control of the instrument temperature is ensured by the IRT using heaters (opened or closed loop). Concerning the power dissipation at CF1 and CF2, it is expected to be in the range 1.9--2.8 and 2.8--6.7 W, respectively. 

Figure \ref{fig:cf} depicts the overall electrical architecture of the IRT Instrument. It includes a warm electrical sub-assembly (DHU) and cold electrical sub-systems: the IRT--FPA, the IRT--WA and the IRT--CUA. The IRT DHU (Data Handling Unit) will have digital interfaces with the S/C and the IRT--FPA, and analog interfaces with the FPA (bias and signals), with the WA (motor control) and with the CUA. A primary power interface with the S/C will complete the electrical interfaces.

\begin{figure} [ht]
   \begin{center}
   \begin{tabular}{c} 
   \includegraphics[height=5.8cm]{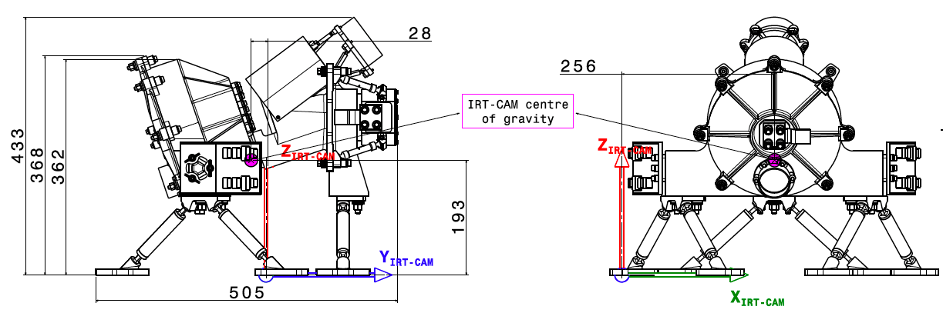}
   \end{tabular}
   \end{center}
   \caption[example] 
   { \label{fig:mec} 
IRT Instrument main dimensions.}
   \end{figure}

The main dimensions of the IRT Camera are depicted in Figure \ref{fig:mec}. The mechanical interface to the optical bench is a hexapod, each of its six feet is bolted with three screws. The IRT DHU main dimensions are 210 mm x 228 mm x 220 mm (not shown). 

Given the design and interfaces presented above the total mass and power budget for the IRT, excluding harnesses, and including typical phase A margins is reported in Table \ref{tab:resources}, where we also report the telemetry budget. The latter is based on the fact, that a series of images in different filters are needed to acquire the GRB afterglow position and the photometric redshift (see later). For a complete sequence up to the photometric redshift we estimate that at least 252 Mbytes of data need to be transmitted to ground. If we include the GRB characterization mode between 25 Mbytes and 1.2 GBytes per GRB are needed. For the GO mode, a minimal amount of 80 kbits/s are needed.

\begin{table}[ht]
\caption{IRT--CAM resources.} 
\label{tab:resources}
\begin{center}       
\begin{tabular}{|c|c|c|c|} 
\hline
Element & Mass (kg) & Average Power (W) &  Maximum/Average \\
&&&Required Telemetry (Gb/day) \\
\hline
IRT--CAM & 32.5 & 36.2 & \\
\hline
IRT-DHU & 5.6 & 30.1  & \\
\hline
Total Instrument & 38.2 & 66.3 & 40/13.5\\
\hline
\end{tabular}
\end{center}
\end{table} 

\section{Operations and in-flight calibration}

\begin{figure} [ht]
   \begin{center}
   \begin{tabular}{c} 
   \includegraphics[height=8.5cm]{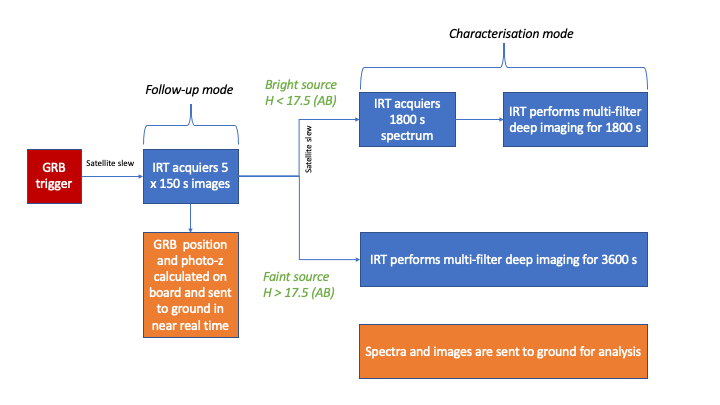}
   \end{tabular}
   \end{center}
   \caption[example] 
   { \label{fig:follow-up} 
IRT science operations after a GRB trigger.}
   \end{figure}

The IRT operation sequence after a GRB trigger is summarized in Figure \ref{fig:follow-up}. Once a GRB is detected by the SXI and/or the XGIS, a slew is requested to the platform in order to place the GRB error box within the IRT photometric FOV. Then, when the S\/C is stabilized, the IRT enters the follow-up mode, where a 150 s exposure for each of the available filters (I, Z, Y, J and H) is acquired; the depth of each image shall exceed 20.4 mag (AB). Thanks to these images and an on-board catalogue (based on Gaia and Euclid surveys), the IRT shall be able to autonomously identify the GRB afterglow candidate, compute its coordinates (to a better than 5 arc seconds accuracy) and its photometric redshift (expected accuracy better than 10\%) and send this information to ground. Then, as a function of the  identified source flux, the IRT enters either the characterization mode, which includes spectroscopy mode for 1800 s (if the source is brighter than 17.5 mag (H, AB)) followed by 1800 of deep imaging, or directly the deep imaging mode for 3600 s. In order to activate the spectroscopic mode, the satellite needs to perform a small slew to put the afterglow positions within the 2$\times$2 arc min spectroscopic FoV. 

Except from GRB follow-up, the IRT can be operated in GO mode: in this mode, in order to cope with the allocated telemetry resources, only three windows of 200 $\times$ 200 pixels centred on three sources (one is the target, the other two astrometric and photometric references) and not the entire FoV frames will be transmitted to ground. IRT can also be operated in calibration mode, see below..

Calibration observations will be performed at regular intervals (the frequency depends on the nature of the information required) to obtain the data needed to produce the calibration files and to monitor the good health of the instrument. To monitor the detector health (bad pixels, flat field, linearity) an on-board calibration unit (CUA) will be used. On the other hand, in order to be compliant with the photometric accuracy requirements ($<$5\%), every 4-6 months calibration observations of known stars will be required. The same is valid for spectrophotometry, where known stars need to be observed over the entire (spectroscopic) FoV. Concerning the wavelength calibration, planetary nebulae with clear emission lines will be observed at different positions on the detector for adjustment and verification of the line centroids. On the launch date of Theseus, once more, it will be possible to use the catalogue produced for Euclid's NISP. 

\section{IRT Expected Performance}

The main IRT characteristics in terms of expected scientific performances is summarized in Table \ref{tab:perfo}.

\begin{table}[ht]
\caption{IRT Performance.} 
\label{tab:perfo}
\begin{center}       
\begin{tabular}{|c|c|} 
\hline
IRT Characteristic & Value \\
\hline
Photometric sensitivity per filter  & I: 20.9 (goal:21.3)\\
(AB, in 150 s, SNR = 5) & Z: 20.7 (goal: 21.2)\\
& Y: 20.4 (goal: 20.8)\\
& J: 20.7 (goal: 21.1)\\
& H: 20.8 (goal: 21.1)\\
\hline
Photometric accuracy & 5\%\\
\hline
Expected photo-z 	accuracy & $<$10\%\\
\hline
Astrometric accuracy & $<$ 5 arc sec in near-real time\\
& $<$ 1 arc sec after ground processing\\
\hline
Spectral resolution & $\geq$400\\
\hline
Spectral sensitivity & 17.5 (goal 19)\\
(AB, H filter, 1800 s, SNR=3 for each spectral bin)  &\\
\hline
\end{tabular}
\end{center}
\end{table} 

In order to precisely estimate the photometric capabilities of the IRT, we have developed a simulation tool, that takes into account the detector performance (readout noise, dark current, pixel size, …), the operating conditions (temperature, photometric aperture), the satellite planned pointing (expected zodiacal light background and out of field stray-light), the observation conditions (source characteristics, exposure and individual frame duration), the filter characteristics, and the optical imperfections of the system. The latter include the satellite jitter and drift, which together with the detector readout noise are the main limitations to the instrument sensitivity. 

\begin{figure} [ht]
   \begin{center}
   \begin{tabular}{c} 
   \includegraphics[height=8.5cm]{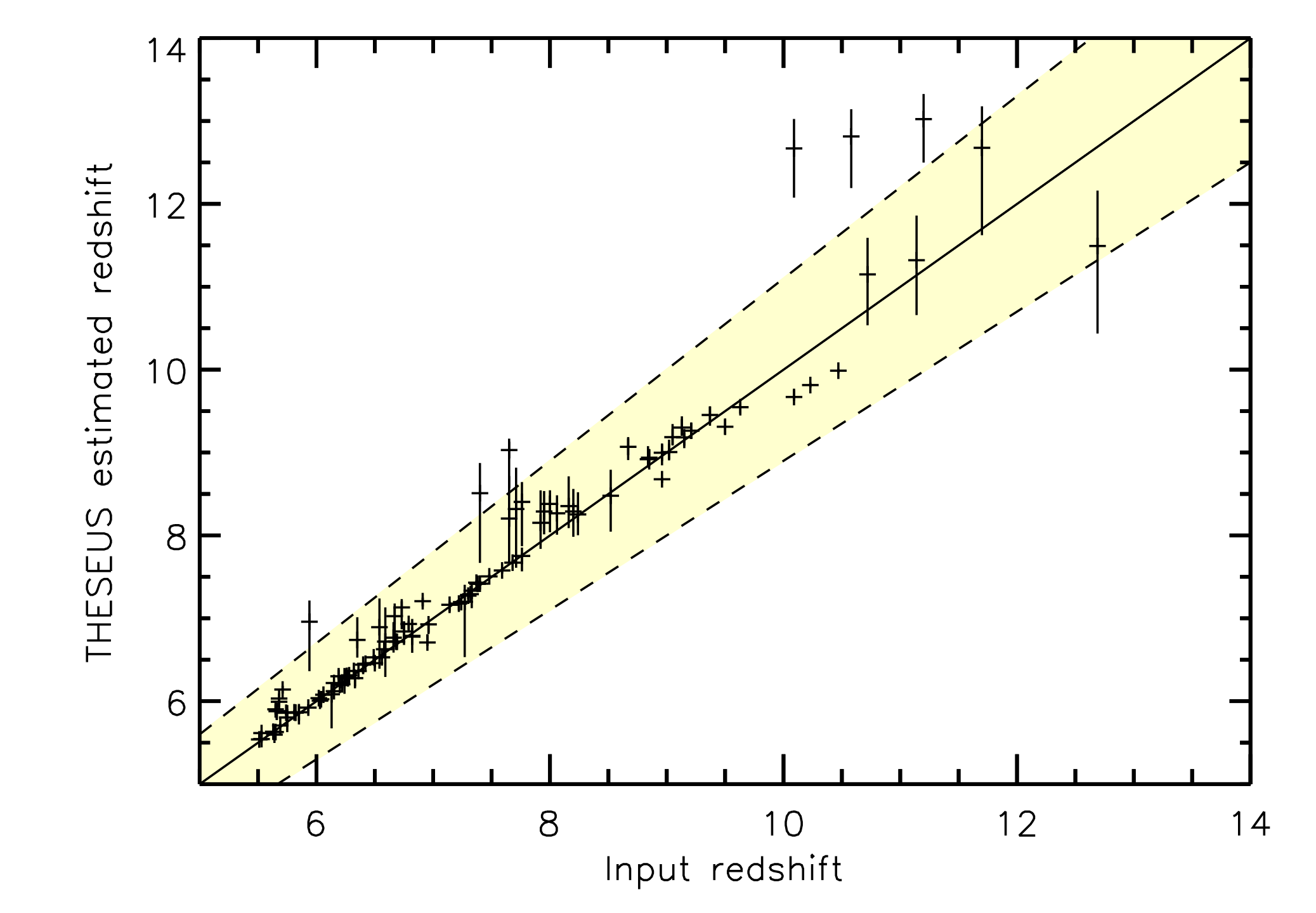}
   \end{tabular}
   \end{center}
   \caption[example] 
   { \label{fig:perf} 
Reconstructed photometric redshift vs. injected ones for a Monte Carlo simulation of the IRT detected GRBs at the end of the photo-z sequence (based on \cite{corre2018}). The yellow stripe represents the 10\% accuracy region.}
   \end{figure}

Taking all these effects into account, we can realistically estimate the expected limiting sensitivity of the IRT. Our simulations indicate that we are always compliant with the requirements of Table \ref{tab:perfo}, both for the ground based (including fine astrometric corrections) and for the on-board near real time data processing algorithms.
In addition, some extra margin can be gained by the optimization of the filters bandpass. With these sensitivities we expect to be able to determine the photometric redshift (through the determination of the Lyman-alpha break wavelength) to an accuracy of better than 10\% in more than 90\% of the observed GRBs, see Figure \ref{fig:perf}. Due to the detector passband we are mainly limited to redshifts in the range 5.5-12.

Detailed simulations have been performed also for the IRT spectroscopic mode.
The following steps have been implemented to provide a preliminary evaluation of the performances of the IRT spectrometric mode:

\begin{enumerate}
\item A preliminary analysis based on a simplified geometric model gives access to the linear dispersion that has to be implemented in order to reach the specified Resolving powers (R). This has been performed at optical design level.
\item Once dispersion is set, we simulate parts of the spectra in a process involving, dispersion, geometrical aspects, diffraction, platform stability, straylight/backgrounds, etc. This analysis gives access to the pixel dependent Instrument Spread Function (ISF), which gives the spectral content of the signal susceptible to be collected by each pixel. Then, the per-pixel noise can be computed to finally derive the SNR. This is done for a few columns of pixels (typically 10 to 20) distributed over the spectral axis, and a few contiguous pixels in the spatial axis (typically 3).
\item Derive the actual resolving power.
\item Derive the processed SNR/R, i.e. after summing the contiguous rows of pixels, and stacking individual exposures done during a sequence
\end{enumerate}

\begin{figure} [ht]
   \begin{center}
   \begin{tabular}{c} 
   \includegraphics[height=8.5cm]{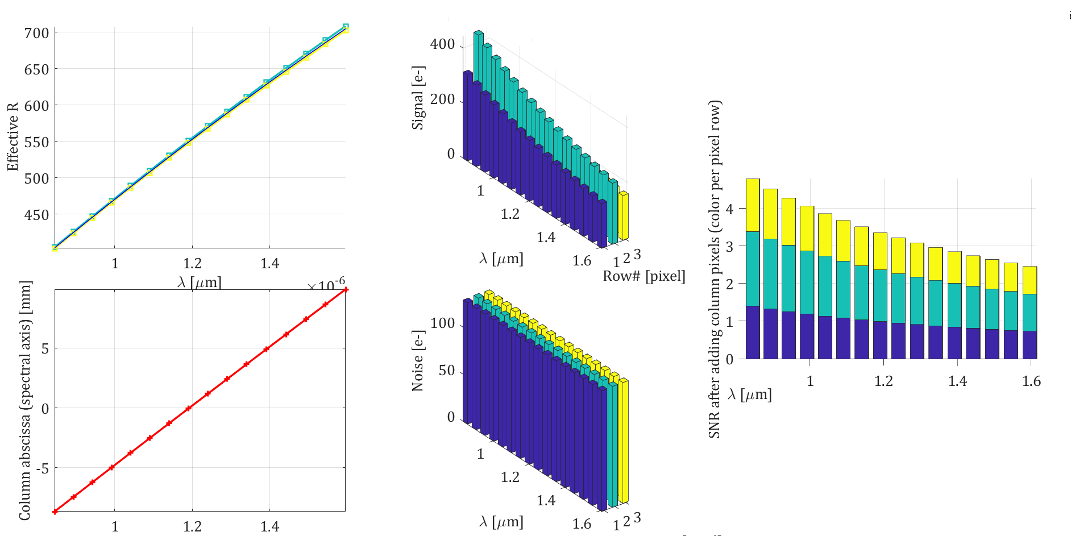}
   \end{tabular}
   \end{center}
   \caption[example] 
   { \label{fig:spe} 
Top left: Actual resolving power as a function of wavelength. Bottom left: Linear dispersion in physical units on the detector plane: spectrum spreads almost linearly over 18.6 mm.
Right panel: Source signal and background signal over three columns and combined SNR with the contribution of each column color coded.}
   \end{figure}

A typical output of these steps is shown in Fig. \ref{fig:spe}.
   
Given the longer exposure of 1800 s (with individual frames of 60 s), the satellite stability plays a strong role in the spectral performance. In order to assess the IRT spectroscopic performance in the most realistic way, we used several S/C simulated drift patterns, expected for the THESEUS orbital conditions and the expected attitude control system performance. We also weighted our results in terms of expected zodiacal background over the foreseen THESEUS sky pointings. The source and background signals have been estimated for each detector pixel, and some margin has been taken for imperfect stacking of the single frames (note that in order to stack the individual 60 s frames we used the “0th” spectral order as a guide). We could in this way estimate the expected SNR of each spectral bin (summed over three detector lines), and this as a function of the resolving power for each wavelength.  With this performance we expect to be able to determine the spectroscopic redshift of GRBs (also for those at z $<$ 5.5), and, for the brightest ones, to determine the neutral absorption column density and to detect the presence of metals.

\section{Conclusions}

We presented the phase A design of the IRT Telescope, which is part of the payload of the THESEUS M5 project. We have shown that we the current design we are able to meet the required scientific performance, and hence, if THESEUS is selected, IRT will play a key role in detecting, localizing and measuring the distance of cosmological Gamma-Ray Bursts in providing essential information to all those facilities interested in the deep Universe science the '30s.

\section*{Acknowledgments}
This work has been partially funded by the French Space Agency (CNES). D.G. acknowledges financial support by LabEx UnivEarthS (ANR-10-LABX-0023 and ANR-18-IDEX-0001).

%
%

\bibliography{report} 
\bibliographystyle{spiebib} 

\end{document}